\documentclass[10pt]{article}
\usepackage{epsfig}
\usepackage{epstopdf}
\usepackage{amsfonts}
\usepackage{amsmath}
\usepackage{amssymb}
\usepackage{graphicx}
\usepackage[colorlinks=true,linkcolor=red,citecolor=blue]{hyperref}
\begin{document}
\title{Branching Fractions  and Direct $CP$ Asymmetries of
$\overline B_s ^0 \to K^0 h^+h^{\prime -}(h^{(\prime)}=K,\pi)$ Decays}
\author{Ying  Li\footnote{Email:liying@ytu.edu.cn}
 \\
{\it  Department of Physics, Yantai University, Yantai 264-005,
China} } \maketitle

Motivated by the recent LHCb collaboration measurements of charmless
three-body decays of $\overline B_s^0$ meson, we calculate the
branching fractions of $\overline B_s ^0 \to K^0 \pi^+\pi^-$,
$\overline B_s ^0 \to K^0 K^+K^-$, $\overline B_s ^0 \to K^0
\pi^+K^-$ and $\overline B_s ^0 \to \overline K^0 K^+\pi^-$ decay
modes using the factorization approach. Both the resonant and
nonresonant contributions are studied in detail. For the decays
$\overline B_s ^0 \to K^0 \pi^+\pi^-$ and $\overline B_s ^0 \to K^0
K^+K^-$, our results agree well with experimental data, and the
former is dominated by the $K^*$, while the latter one is dominated
by the nonresonant contribution.  Considering the flavor $SU(3)$
symmetry violation, the sum of branching fractions of $\overline B_s
^0 \to K^0 \pi^+K^-$ and $\overline B_s ^0 \to \overline K^0
K^+\pi^-$ could accommodate the data well too. It should be noted
that both branching fractions are sensitive to the scalar density
$\langle K\pi| \bar s q|0\rangle$. Furthermore, the resonant
contributions are dominated by the scalar $K_0^*(1430)$. We hope
that these branching fractions could be measured individually in the
experiments so as to test the factorization approach and the flavor
$SU(3)$ asymmetry.  Moreover, the  direct $CP$ asymmetries of these
decays are also investigated, which could be measured in the running
LHCb experiment and Super-b factory in the future. \\
\\
{\bf Key Words:}   $B_s^0$ three-body decay, Factorization Approach,
CP
violation\\
\\
{\bf PACS:}13.25.Hw, 11.30.Er, 14.40.Nd

\newpage
\section{Introduction}\label{S1}
In the recent years, the charmless three-body decays of $B$ mesons
have attached a great deal of attention, because by studying them
one can determine the Cabibbo-Kabayashi-Maskawa (CKM) parameters or
search for the possible new physics effect beyond the standard
model. For example, the Dalitz-plot analysis combined with flavor
$\mathrm{SU}(3)$ symmetry allows us to extract the angle $\gamma$
cleanly from $B \to K \pi\pi$ and $B \to KKK$ decays
\cite{Ciuchini:2006kv,Gronau:2006qn,ReyLeLorier:2011ww}. However,
the three-body decays of $B$ mesons are more complicated than the
two-body cases, because both resonant (vector or scalar) and
nonresonant contributions involve the hadronic matrix elements. The
interference between resonant and nonresonant amplitudes makes it
rather hard to disentangle these distinct contributions and extract
the nonresonant one, so it is very difficult to measure the direct
three-body decays experimentally. Over the recent years, thanks to
the two $B$ factories and LHCb experiment, remarkable progress in
measuring the branching fractions and direct $CP$ asymmetries of the
three-body decays has been made by using the Dalitz-plot analysis
(for a review see ref. \cite{Amhis:2012bh}).

On the theoretical side, the charmless three-body decays of heavy
mesons have been studied within the different approaches, such as
the factorization approach (FA) \cite{Deshpande:1995nu,
Fajfer:1998yc, Bajc:1998bs,Fajfer:2004cx,Cheng:2002qu,
Cheng:2005bg,Cheng:2005ug,Cheng:2007si,Cheng:2013dua}, diagrammatic
approach combined with $\mathrm{SU}(3)$ symmetry
\cite{Gronau:2003ep, Gronau:2005ax, Lorier:2010xf,
Bhattacharya:2013cvn}, perturbative QCD approach
\cite{Chen:2002th,Wang:2014ira}, and other approaches
\cite{Zhang:2013oqa, Xu:2013dta, Xu:2013rua}. FA, based on the
phenomenological factorization model, has been applied in
calculating three-body decays of heavy meson widely, although
factorization has not been proved in the three-body decays. Within
the FA, most predicted branching fractions and direct $CP$
asymmetries of $B \to PPP$ decays \cite{Cheng:2002qu, Cheng:2005bg,
Cheng:2005ug, Cheng:2007si, Cheng:2013dua} agree with the
experimental data well, except for decay $\overline B^0 \to
K^+K^-\pi^0$.

Here, we will review the FA  briefly by taking $B^-\to
\pi^+\pi^-\pi^-$ as an example. Under the FA, the amplitude of decay
$B^-\to \pi^+\pi^-\pi^-$ is usually split into three distinct
factorizable terms: (i) the current-induced process with a meson
emission, $\langle B^-\to \pi^+\pi^-\rangle\times \langle 0\to
\pi^-\rangle$, (ii) the transition process, $\langle B^-\to
\pi^-\rangle\times \langle 0\to \pi^+\pi^-\rangle$, and (iii) the
annihilation process $\langle B^-\to 0\rangle\times \langle 0\to
\pi^+\pi^-\pi^-\rangle$, where $\langle A\to B\rangle$ stands for an
$A\to B$ transition matrix element. One of the nonresonant
contributions due to $\langle B^-\to \pi^+\pi^-\rangle$ has been
studied on the basis of the heavy meson chiral perturbative theory
(HMChPT) \cite{Yan:1992gz, Wise:1992hn, Burdman, Lee:1992ih},
although applicability of this framework in the whole kinematics
region is still controversial \cite{wei:2013}. However, it could
lead to large branching fraction (${\cal O}(10^{-5})$)
\cite{Deshpande:1995nu, Fajfer:1998yc}, which disagrees with the
experimental data ($5.3 \times10^{-6} $) from BaBar
\cite{Aubert:2009av}. In fact, this issue can be understood
considering the applicability of the HMChPT. When the HMChPT is
applied to three-body decays, two of the final-state pseudoscalars
should be soft. If the soft meson result is assumed to be the same
in the whole Dalitz plot, the decay rate will be greatly
overestimated. To overcome this issue, Cheng {\it et al.} proposed
in refs \cite{Cheng:2005bg,Cheng:2005ug,Cheng:2007si,Cheng:2013dua}
to parameterize the momentum dependence of nonresonant amplitudes
$\langle B \to  PP\rangle$ in an exponential form $e^{-\alpha_{_{\rm
NR}}p_B\cdot(p_i+p_j)}$ so that the HMChPT results are recovered in
the soft pseudoscalar meson limit. The tree-dominated
$B^-\to\pi^+\pi^-\pi^-$ decay data is used to fix the unknown
parameter $\alpha_{_{\rm NR}}$. Besides from the current-induced
process, the matrix elements $\langle \pi^+\pi^-|\bar q\gamma_\mu
q|0\rangle$ and $\langle \pi^+\pi^-|\bar d d|0\rangle$ also receive
nonresonant contributions. In principle, the weak vector form factor
of the former matrix element can be related to the charged pion
electromagnetic (e.m.) form factors. However, unlike the kaon case,
the time-like e.m. form factors of the pions are not measured well
enough allowing us to determine the nonresonant parts. Therefore,
the nonresonant contribution to $\langle \pi^+\pi^-|\bar q\gamma_\mu
q|0\rangle$ is always ignored.  The matrix element $\langle
\pi^+\pi^-|\bar d d|0\rangle$ is related to $\langle K^+K^-|\bar
ss|0\rangle$ via SU(3) flavor symmetry. As for the resonant
contributions to three-body decays, vector and scalar resonances
contribute to the two-body matrix elements $\langle
P_1P_2|V_\mu|0\rangle$ and $\langle P_1P_2|S|0\rangle$,
respectively. They can also contribute to the three-body matrix
element $\langle P_1P_2|V_\mu-A_\mu|0\rangle$. Resonant effects are
described in terms of the usual Breit-Wigner formalism. In this
manner, the relevant resonances which contribute to the 3-body
decays of interest could be figured out. In conjunction with the
nonresonant contribution, the total rates for three-body decays are
well calculated.

Very recently, corresponding to an integrated luminosity of
$1.0~\mathrm{fb}^{-1}$ recorded at a centre-of-mass energy of 7 TeV,
LHCb collaboration published their first measurements of the
branching fractions of three-body decays of  $B_s^0$ meson
\cite{Aaij:2013uta} as follows:
\begin{eqnarray}
Br{(B_s^0 \to K^0\pi^+\pi^-)} &=& (14.3 \pm 2.8 \pm 1.8 \pm 0.6)\times10^{-6} \,,\\
Br{(B_s^0 \to K^0K^\pm\pi^\mp)}  &=& (73.6 \pm 5.7 \pm 6.9 \pm 3.0)\times10^{-6} \,,\\
Br{(B_s^0 \to K^0K^+K^-)}   &\in& [0.2;3.4] \times10^{-6}
\; {\rm at \;\; 90\% \; CL} \,.
\end{eqnarray}
Since these decays have never been explored before, we will
calculate the branching fractions in this work using the FA proposed
by Cheng {\it et.al.} so as to test FA in $\overline B^0_s$ decays.
The resonant and nonresonant contributions of these decays will be
studied, which are important in measuring the branching fractions of
$\overline B^0_s \to K V$ and $\overline B^0_s \to K S$
experimentally. Furthermore, we will calculate the $CP$ asymmetries
of these decays, which may be helpful to extract the CKM angle
$\gamma$.  All results could be checked in the current LHCb
experiment and Super-b factory in the future.

In the following work, we will systematically use the FA to
calculate the $\overline B_s ^0 \to K^0 h^+h^{\prime -}$  and
present the formulas in Sec.\ref{S2}.  The numerical results and
some discussions are given in Sec. \ref{S3} . We will summarize this
work in Sec. \ref{S4} lastly.

\section{Analytic Formalism}\label{S2}
\subsection{The Effective Hamiltonian}
Under the factorization hypothesis, the matrix elements of the decay
amplitudes are given by
\begin{eqnarray} \label{eq:factamp}
\langle P_1P_2P_3|{\cal H}_{\rm eff}| \overline B_s ^0 \rangle
 =\frac{G_F}{\sqrt2}\sum_{p=u,c}\lambda_p^{(r)} \langle P_1P_2P_3|T_p^{(r)}| \overline B_s ^0 \rangle,
\end{eqnarray}
where $\lambda_p^{(r)}\equiv V_{pb} V^*_{pr}$ with $r=d,s$.  For
$K\pi\pi$ and $KKK$ modes, $r=d$; and for $KK\pi$ channels, $r=s$.
The Hamiltonian $T_p^{(r)}$ has the expression \cite{Buras}
\begin{eqnarray} \label{eq:Tp}
 T_p^{(r)}&=&
 a_1 \delta_{pu} (\bar u b)_{V-A}\otimes(\bar r u)_{V-A}
 +a_2 \delta_{pu} (\bar r b)_{V-A}\otimes(\bar u u)_{V-A}
 +a_3(\bar r b)_{V-A}\otimes\sum_q(\bar q q)_{V-A}
 \nonumber\\
 &&+a^p_4\sum_q(\bar q b)_{V-A}\otimes(\bar r q)_{V-A}
   +a_5(\bar r b)_{V-A}\otimes\sum_q(\bar q q)_{V+A}
       \nonumber\\
 &&-2 a^p_6\sum_q(\bar q b)_{S-P}\otimes(\bar r q)_{S+P}
 +a_7(\bar r b)_{V-A}\otimes\sum_q\frac{3}{2} e_q (\bar q q)_{V+A}
\nonumber\\
 &&-2a^p_8\sum_q(\bar q b)_{S-P}\otimes\frac{3}{2} e_q
             (\bar r q)_{S+P}
 +a_9(\bar r b)_{V-A}\otimes\sum_q\frac{3}{2}e_q (\bar q q)_{V-A}\nonumber\\
 &&+a^p_{10}\sum_q(\bar q b)_{V-A}\otimes\frac{3}{2}e_q(\bar r
 q)_{V-A},
\end{eqnarray} with $(\bar q q')_{V\pm A}\equiv \bar
q\gamma_\mu(1\pm\gamma_5) q'$, $(\bar q q')_{S\pm P}\equiv\bar
q(1\pm\gamma_5) q'$ and a summation over $q=u,d,s$ being implied.
For the effective Wilson coefficients at the renormalization scale
$\mu=2.1$~GeV, we shall follow \cite{Cheng:2007si} and use
\begin{eqnarray} \label{eq:ai}
 && a_1\approx0.99+0.037 i,\quad a_2\approx 0.19-0.11i, \quad a_3\approx -0.002+0.004i, \quad a_5\approx
 0.0054-0.005i,  \nonumber \\
 && a_4^u\approx -0.03-0.02i, \quad a_4^c\approx
 -0.04-0.008i,\quad
 a_6^u\approx -0.06-0.02i, \quad a_6^c\approx -0.06-0.006i,
 \nonumber\\
 &&a_7\approx 0.54\times 10^{-4} i,\quad a_8^u\approx (4.5-0.5i)\times
 10^{-4},\quad
 a_8^c\approx (4.4-0.3i)\times
 10^{-4},   \nonumber\\
 && a_9\approx -0.010-0.0002i,\quad
 a_{10}^u \approx (-58.3+ 86.1 i)\times10^{-5},\quad
 a_{10}^c \approx (-60.3 + 88.8 i)\times10^{-5},
\end{eqnarray}
In the above coefficients, the strong phases are from vertex
corrections and penguin contractions, which have  been calculated
within the QCD factorization approach \cite{BBNS}.

\subsection{$\overline B_s^0\to K^0\pi^+\pi^-$}
With the effective Hamiltonian and the equation of motion, we obtain
the $\overline B_s^0\to K^0\pi^+\pi^-$ decay amplitude as
\begin{eqnarray} \label{eq:AKpipi}
\langle K^0 \pi^+ \pi^-|T_p|\overline B_s^0\rangle &=& \langle
K^0\pi^+ |(\bar u b)_{V-A}|\overline B_s^0\rangle \langle
\pi^-|(\bar d u)_{V-A}|0\rangle
 \left[a_1 \delta_{pu}+a^p_4+a_{10}^p- r_\chi^\pi(a^p_6+a^p_8)\right]
  \nonumber\\
&& + \langle \pi^+ \pi^- |(\bar s b)_{V-A}|\overline B_s^0\rangle \langle
\pi^-|(\bar d s)_{V-A}|0\rangle
 \left[a^p_4-\frac{1}{2}a_{10}^p- r_\chi^K(a^p_6-\frac{1}{2}a^p_8)\right] \nonumber \\
&& + \langle K^0|(\bar db)_{V-A}|\overline B_s^0\rangle\langle\pi^+\pi^-|(\bar
uu)_{V-A}|0\rangle\Big[a_2\delta_{pu}+a_3+a_5+a_7+a_9\Big] \nonumber \\
&& + \langle K^0|(\bar db)_{V-A}|\overline B_s^0\rangle\langle\pi^+\pi^-|(\bar
dd)_{V-A}|0\rangle\Big[a_3+a_4^p+a_5-{1\over 2}(a_7+a_9+a_{10}^p)\Big] \nonumber \\
&& + \langle K^0|(\bar db)_{V-A}|\overline B_s^0\rangle\langle\pi^+\pi^-|(\bar
ss)_{V-A}|0\rangle\Big[a_3+a_5-{1\over 2}(a_7+a_9)\Big] \nonumber \\
&& +\langle K^0|\bar db|\overline B_s^0\rangle\langle\pi^+\pi^-|\bar dd|0\rangle\Big[-2a_6^p+a_8^p\Big] \nonumber \\
&& +\langle K^0 \pi^+ \pi^-|(\bar sd)_{V-A}|0\rangle \langle0|(\bar sb)_{V-A}|\overline B_s^0\rangle
\Big[a_4^p-{1\over 2}a_{10}^p\Big] \nonumber \\
&& +\langle K^0 \pi^+ \pi^-|\bar s(1+\gamma_5)d|0\rangle\langle0|\bar s\gamma_5b|\overline
B_s^0\rangle\Big[2a_6^p-a_8^p\Big],
\end{eqnarray}
where $r_\chi^\pi(\mu)={2m_\pi^2\over m_b(\mu)(m_d(\mu)-m_u(\mu))}$.
It should be noted that  $\langle \pi^+\pi^-|(\bar
dd)_{V-A}|0\rangle=-\langle \pi^+\pi^-|(\bar uu)_{V-A}|0\rangle$
because of isospin symmetry. Besides, the matrix element $\langle
\pi^+\pi^-|(\bar ss)_{V-A}|0\rangle$ is suppressed heavily by the
Okubo-Zweig-Iizuka (OZI) rule. Moreover, there exist two weak
annihilation contributions, where the $\overline B_s^0$ meson is
annihilated  into vacuum  and a final state with three mesons is
then created,  as the last two term are shown in the above equation.
However, from the results of $B \to PPP$ decays, the contributions
from annihilations are fairly small because of  power and $\alpha_s$
suppressions, so we will ignore them in the numerical calculations
in the current work. \footnote{In  the chiral limit, $\langle K^0
\pi^+ \pi^-|(\bar sd)_{V-A}|0\rangle$ has been proven to be zero
\cite{Cheng:2002qu}. For the term $\langle K^0 \pi^+ \pi^-|\bar
s(1+\gamma_5)d|0\rangle$, it is penguin induced and power
suppressed. Thus, its contribution could be dropped safely.}

For the current-induced process, the three-body matrix element
$\langle K^0\pi^+ |(\bar u b)_{V-A}|\overline B_s^0\rangle$ could be
parameterized as \cite{Lee:1992ih}
\begin{eqnarray} \label{eq:romegah}
 \langle K^0(p_1)\pi^+ (p_2) |(\bar u b)_{V-A}|\overline B_s^0(p_B)\rangle
 &=&i r (p_B-p_1-p_2)_\mu+i\omega_+(p_2+p_1)_\mu+i\omega_-(p_2-p_1)_\mu \nonumber\\
 &&+h\,\epsilon_{\mu\nu\alpha\beta}p_B^\nu (p_2+p_1)^\alpha(p_2-p_1)^\beta.
\end{eqnarray}
The form factors $\omega_\pm$ and $r$ have the expressions as \cite{Lee:1992ih}
\begin{eqnarray}\label{eq:r&omega}
\omega_+ &=& -{g\over f_\pi
f_K}\,{f_{B^*}m_{B^*}\sqrt{m_{B_s}m_{B^*}}\over
 s_{23}-m_{B^*}^2}\left[1-{(p_B-p_1)\cdot p_1\over
 m_{B^*}^2}\right]+{f_{B_s}\over 2f_\pi f_K},   \nonumber \\
 \omega_- &=& {g\over f_\pi f_K}\,{f_{B^*}m_{B^*}\sqrt{m_{B_s}m_{B^*}}\over
 s_{23}-m_{B^*}^2}\left[1+{(p_B-p_1)\cdot p_1\over
 m_{B^*}^2}\right], \nonumber  \\
 r &=& {f_{B_s}\over 2f_\pi f_K}-{f_{B_s}\over
 f_\pi f_K}\,{p_B\cdot(p_2-p_1)\over
 (p_B-p_1-p_2)^2-m_{B_s}^2}+{2gf_{B^*}\over f_\pi f_K}\sqrt{m_{B_s}\over
 m_{B^*}}\,{(p_B-p_1)\cdot p_1\over s_{23}-m_{B^*}^2} \nonumber \\
 &&- {4g^2f_{B_s}\over f_\pi f_K}\,{m_{B_s}m_{B^*}\over
 (p_B-p_1-p_2)^2-m_{B_s}^2}\,{p_1\!\cdot\!p_2-p_1\!\cdot\!(p_B-p_1)\,p_2\!\cdot\!
 (p_B-p_1)/m_{B^*}^2 \over s_{23}-m_{B^*}^2 },
\end{eqnarray}
where $s_{ij}\equiv (p_i+p_j)^2$. $g$ is a heavy-flavor independent
strong coupling which has been extracted from the CLEO measurement
of the $D^{*+}$ decay width \cite{Ahmed:2001xc},
$|g|=0.59\pm0.01\pm0.07$. In this work, we also follow
\cite{Yan:1992gz} and adopt its sign as negative. Thus, we drive the
current-induced amplitude as:
\begin{eqnarray}
\label{eq:AHMChPT}
 &&A_{\rm current-ind}\equiv\langle \pi^-(p_3)|(\bar du)_{V-A}|0\rangle
 \langle  K^0(p_1)\pi^+ (p_2)|(\bar u b)_{V-A}|B^-\rangle \nonumber\\
 && =-\frac{f_\pi}{2}\left[2 m_3^2 r+(m_B^2-s_{12}-m_3^2) \omega_+ +(s_{23}-s_{13}-m_2^2+m_1^2)
 \omega_-\right]\,e^{-\alpha_{NR}
p_B\cdot(p_{1}+p_{2})}e^{i\phi_{12}}.
\end{eqnarray}
As stated in Sec.\ref{S1}, the exponential form $e^{-\alpha_{\rm NR}
p_B\cdot(p_{1}+p_{2})}$ is introduced so that the HMChPT results are
recovered in the soft meson region and
\begin{eqnarray} \label{eq:alphaNR}
 \alpha_{\rm NR}=0.081^{+0.015}_{-0.009}\,{\rm GeV}^{-2},
\end{eqnarray}
which is constrained from the tree dominated decay  $B^-\to
\pi^+\pi^-\pi^-$ \footnote{In the above calculations, the
heavy-quark chiral effective approach has been adopted, where the
light  pseudoscalar mesons  are regarded as Goldstone  bosons. Thus,
the $SU(3)$ symmetry breaking effects in  $\alpha_{\rm NR}$ have not
been invovled for their negligible uncertainties.}. The unknown
strong phase $\phi_{12}$ is set to be zero for simplicity.

In this decay mode, vector meson ($K^*$) and scalar resonances
($K_{0}^*(1430)$) also contribute to the three-body matrix element
$\langle K^0(p_1)\pi^+ (p_2) |(\bar u b)_{V-A}|\overline
B_s^0(p_B)\rangle$, whose effects are described in terms of the
usual Breit-Wigner formalism. So, we have the expression as
\begin{eqnarray} \label{eq:m.e.pole}
 \langle K^0(p_1)\pi^+(p_2)|(\bar ub)_{V-A}|\overline B_s^0\rangle^R
& =& {g^{K^{*+}\to K^0\pi^+}\over s_{12}-m_{K^{*+}}^2+im_{K^{*+}}\Gamma_{K^{*+}}}\sum_{\rm pol}
\varepsilon^*\cdot (p_1-p_2)\langle K^{*+}|(\bar
ub)_{V-A}|\overline B_s^0\rangle \nonumber \\
&-&  {g^{{K_0^{*+}}\to K^0\pi^+}\over s_{12}- m_{K_0^{*+}}^2+im_{K_0^{*+}}\Gamma_{K_0^{*+}}}\langle K_0^{*+}|(\bar
ub)_{V-A}|\overline B_s^0\rangle,
 \end{eqnarray}
where we have ignored the contribution of $K^*(1410), K^*(1680),\cdots$. Hence,
\begin{eqnarray}\label{3formfactor}
 && \langle K^0(p_1)\pi^+(p_2)|(\bar ub)_{V-A}|\overline B^0_s\rangle^R ~\langle \pi^-(p_3)|(\bar
du)_{V-A}|0\rangle \nonumber \\
&=&{-f_\pi}\,{g^{K^{*+}\to K^0\pi^+}\over
s_{12}-m_{K^*}^2+im_{K^*}\Gamma_{K^*}}\left\{\Big[s_{13}-s_{23}+\frac{(m_{B_s}^2-m_\pi^2)(m_\pi^2-m_K^2)}
{m_{K^*}^2}\Big]\Big[ m_{K^*}A_0^{B_sK^*}(q^2)\right.\nonumber
\\&&+\frac{A_2^{B_sK^*}(q^2)}{2(m_{B_s}+m_{K^*})}(s_{12}-m_{K^*}^2)
\Big]+\left (m_\pi^2-m_K^2\right)\left (1-\frac{s_{12}}{m_{K^*}^2}
\right)\Big[ m_{K^*}A_0^{B_sK^*}(q^2)
\nonumber\\&&\left.-(m_{B_s}+m_{K^*})A_0^{B_sK^*}(q^2)
+\frac{A_2^{B_sK^*}(q^2)}{2(m_{B_s}+m_{K^*})}(s_{12}-m_{K^*}^2)
\Big]\right\}\nonumber\\&& + f_K{g^{K_{0}^*\to K^-\pi^+}\over
s_{12}-
m_{K_{0}^*}^2+im_{K_{0}^*}\Gamma_{K_{0}^*}}\Big[(m_{B_s}^2-m_{K_{0}^*}^2)F_0^{B_sK_0^*}(q^2)
+(m_{K_{0}^*}^2-s_{12})F_1^{B_sK_0^*}(q^2)\Big].
\end{eqnarray}
with $q^2=(p_B-p_1-p_2)^2=p_3^2$. In the above formulaes, the
definitions of decay constants and form factors are referred to
Refs. \cite{Cheng:2013dua,Wirbel:1985ji,Cheng:2003sm}.

For the transition processes that are penguin induced or color
suppressed, because the time-like e.m. form factors of two pions
have not been measured well enough, we will thus ignore the
nonresonant contributions and only consider the contributions from
the vector and scalar mesons. Hence, the amplitude of the transition
process is read as
\begin{eqnarray}
&& \langle \pi^+(p_2)\pi^-(p_3)|(\bar uu)_{V-A}|0\rangle^R \langle
K^0(p_1)|(\bar d b)_{V-A}|\overline B_s^0\rangle
= -F_1^{B_sK}(s_{23})F^{\pi^+\pi^-}_R(s_{23})\left(s_{12}-s_{13}\right), \nonumber \\
&& \langle \pi^+(p_2)\pi^-(p_3)| \bar dd|0\rangle^R \langle
K^0(p_1)| \bar db|\overline B_s^0\rangle= - {m_{B_s}^2-m_K^2\over
m_b-m_d} F_0^{B_sK}(s_{23})\sum_{i}\frac{m_{{f_0}_i} \bar
f^d_{{f_0}_i} g^{{f_0}_i\to \pi^+\pi^-}}{s_{23}-m_{{f_0}_i}^2+i
 m_{{f_0}_i}\Gamma_{{f_0}_i}},
\end{eqnarray}
with the definition of the form factor $F^{\pi^+\pi^-}_R$:
\begin{eqnarray}
F^{\pi^+\pi^-}_R(s)={1\over\sqrt{2}}\sum_i{m_{\rho_i}f_{\rho_i}g^{\rho_i\to
\pi^+\pi^-}\over s- m_{\rho_i}^2+im_{\rho_i}\Gamma_{\rho_i}},
\end{eqnarray}
where $\rho_i=\rho,\rho(1450),\cdots$ and
$f_0=f_0(980),f_0(1370),f_0(1500),\cdots$. The scalar decay constant
$\bar f_{{f_0}_i}^q$ is defined by $\langle {f_0}_i|\bar q
q|0\rangle=m_{{f_0}_i} \bar f^q_{{f_0}_i}$, and $g^{{f_0}_i\to
\pi^+\pi^-}$ is the strong coupling of the ${f_0}_i\to \pi^+\pi^-$
decay. In the practical numerical calculations, the higher excited
states of vector mesons have been ignored for their negligible
contributions.

For the scalar meson $f_0(980)$, we will consider it as the
conventional $q\bar q$, though the quark structure of the light
scalar mesons below or near 1 GeV has been quite controversial.
Because some experimental evidences indicate that $f_0(980)$ is not
purely an $s\bar s$ state \cite{Beringer:1900zz}, we write the
flavor wave functions of the $f_0(980)$ as:
\begin{eqnarray} \label{eq:fsigmaMix}
 |f_0(980)\rangle = |s\bar s\rangle\cos\theta+|n\bar n\rangle\sin\theta,
\end{eqnarray}
with $n\bar n\equiv (\bar uu+\bar dd)/\sqrt{2}$. Experimental
implications for the mixing angle have been discussed in detail in
\cite{Cheng:2002ai}. By assuming 2-quark bound state  for
$f_0(980)$, the observed large rates of $B \to f_0(980)K$ and
$f_0(980)K^*$ modes can be explained in QCDF with the mixing angle
$\theta$ in the vicinity of $20^\circ$ \cite{Cheng:2005nb}. So, we
use $\theta=20^\circ$ in this work.
\subsection{$\overline B_s^0\to K^0 K^+ K^- $}
The factorizable $\overline B_s^0\to K^0 K^+ K^- $ decay amplitude is given by
\begin{eqnarray} \label{eq:AKKK}
\langle K^0 K^+K^-|T_p|\overline B_s^0\rangle &=& \langle K^+K^-
|(\bar sb)_{V-A}|\overline B_s^0\rangle \langle K^0|(\bar
ds)_{V-A}|0\rangle \left[a^p_4-{1\over
2}a_{10}^p-r_\chi^K(a^p_6-{1\over 2}a^p_8)\right]
 \nonumber\\
&& + \langle K^0|(\bar db)_{V-A}|\overline B_s^0\rangle
\langle K^+K^-|(\bar uu)_{V-A}|0\rangle\Big[a_2\delta_{pu}+a_3+a_5+a_7+a_9\Big] \nonumber \\
&& + \langle K^0|(\bar db)_{V-A}|\overline B_s^0\rangle
\langle K^+K^-|(\bar dd)_{V-A}|0\rangle\Big[a_3+a_4^p+a_5-{1\over 2}(a_7+a_9+a_{10}^p)\Big] \nonumber \\
&& + \langle K^0|(\bar db)_{V-A}|\overline B_s^0\rangle
\langle K^+K^-|(\bar ss)_{V-A}|0\rangle\Big[a_3+a_5-{1\over 2}(a_7+a_9)\Big] \nonumber \\
&&+\langle K^+|(\bar u b)_{V-A}|\overline B_s^0\rangle \langle K^-
K^0|(\bar d u)_{V-A}|0\rangle \Big[a_1\delta_{pu}+a_4^p+a_{10}^p\Big]\nonumber\\
&& +\langle K^0|\bar db|\overline B_s^0\rangle \langle K^+K^-|\bar
dd|0\rangle \Big[-2a_6^p+a_8^p\Big] \nonumber \\
&&+\langle K^+|\bar u b|\overline B_s^0\rangle \langle K^-
K^0|\bar d u|0\rangle \Big[-2a^p_6-2a^p_8\Big]\nonumber\\
&& +\langle K^0 K^+K^-|(\bar ds)_{V-A}|0\rangle \langle0|(\bar
sb)_{V-A}|\overline B_s^0\rangle\Big[a_4^p-{1\over 2}a_{10}^p\Big] \nonumber \\
&& +\langle K^0 K^+K^-|\bar d(1+\gamma_5)s|0\rangle \langle0|\bar
s\gamma_5b|\overline B_s^0\rangle\Big[2a_6^p-a_8^p\Big].
\end{eqnarray}
For the current-induced process with a kaon emission, the form
factors $r$ and $\omega_\pm$ for the three-body matrix element
$\langle K^+K^- |(\bar sb)_{V-A}|\overline B_s^0\rangle $  evaluated
in the framework of HMChPT are similar to that of
Eq.(\ref{eq:r&omega}) except that $f_\pi$ is replaced by $f_K$. This
process also receives the contributions of vector ($\phi$) and
scalar ($f_0$) resonants by
\begin{multline}
\langle K^+(p_2)K^-(p_3)|(\bar sb)_{V-A}|\overline B_s^0\rangle^R ~\langle K^0(p_1)|(\bar
ds)_{V-A}|0\rangle  \\=
 -{f_K\over 2}\,{g^{\phi\to K^+K^-}\over
s_{23}-m_{\phi}^2+im_{\phi}\Gamma_{\phi}}(s_{12}-s_{13})\Big[
(m_{B_s}+m_{\phi})A_1^{B_s\phi}(q^2) - {A_2^{B_s\phi}(q^2)\over
m_{B_s}+m_{\phi}}
(m_{B_s}^2-s_{23})\\-2m_{\phi}[A_3^{B_s\phi}(q^2)-A_0^{B_s\phi}(q^2)]\Big]
+ f_K\sum_i{g^{f_{0i}\to K^+K^-}\over
s_{23}-m_{f_{0i}}^2+im_{f_{0i}}\Gamma_{f_{0i}}}F_0^{B_sf_{0i}^s}(q^2)(m_{B_s}^2-s_{23}).
\end{multline}

For the transition amplitude, in addition to the $b\to u$ tree
transition, we need to consider the nonresonant contributions to the
$b\to s$ penguin amplitude
\begin{eqnarray}
 A_1 &=& \langle K^{0}(p_1)|(\bar d b)_{V-A}|\overline B_s^0\rangle
  \langle K^+(p_2) K^-(p_3)|(\bar qq)_{V-A}|0\rangle,   \\
 A_2 &=& \langle K^+(p_2)|(\bar u b)_{V-A}|\overline B_s^0\rangle
  \langle K^0(p_1) K^-(p_3)|(\bar d u)_{V-A}|0\rangle,  \\
 A_3 &=& \langle K^{0}(p_1)|\bar d b|\overline B_s^0\rangle
       \langle K^+(p_2) K^-(p_3)|\bar dd|0\rangle,  \\
 A_4 &=& \langle K^+(p_2)|\bar u b|\overline B_s^0\rangle
       \langle K^0(p_1) K^-(p_3)|\bar d u|0\rangle.
\end{eqnarray}
We firstly calculate the two-kaon creation matrix element $A_1$,
which could be expressed in terms of the time-like kaon current form
factors as
\begin{eqnarray} \label{eq:KKweakff}
 \langle K^+(p_2) K^-(p_3)|\bar q\gamma_\mu q|0\rangle
= (p_{K^+}-p_{K^-})_\mu F^{K^+K^-}_q.
\end{eqnarray}
The weak vector form factor $F^{K^+K^-}_q$ is related to the kaon
electromagnetic (e.m.) form factors $F^{K^+K^-}_{\rm em}$.
Phenomenologically, the e.m. form factors receive resonant and
nonresonant contributions and can be expressed by
\begin{eqnarray} \label{eq:KKemff}
F^{K^+K^-}_{\rm em}= F^{KK}_\rho+F^{KK}_\omega+F^{KK}_\phi+F_{NR}.
\end{eqnarray}
It follows from Eqs. (\ref{eq:KKweakff}) and (\ref{eq:KKemff}) that
\begin{eqnarray}
  &&F^{K^+K^-}_u=F^{KK}_\rho+3 F^{KK}_\omega+\frac{1}{3}(3F_{NR}-F'_{NR}),
 \nonumber\\
  &&F^{K^+K^-}_d=-F^{KK}_\rho+3 F^{KK}_\omega,
 \nonumber\\
 && F^{K^+K^-}_s=-3 F^{KK}_\phi-\frac{1}{3}(3 F_{NR}+2F'_{NR}),
 \label{eq:FKKisospin}
\end{eqnarray}
where the isospin symmetry has been used. The resonant and
nonresonant terms  can be parameterized as $F_{h}(s_{23})$ and
$F^{(\prime)}_{NR}(s_{23})$, respectively. Since their expressions
have been given explicitly in Refs.~\cite{Cheng:2005bg,
Cheng:2005ug, Cheng:2007si, Cheng:2013dua}, we will not list them
here. With  the equation of motion, we therefore obtain:
\begin{eqnarray}
 A_1 &=& (s_{12}-s_{13}) F_1^{B_sK}(s_{23}) F^{K^+K^-}_q(s_{23}).
 \end{eqnarray}

In $A_3$, although the nonresonant contribution vanishes as both
$K^+$ and $K^-$ do not contain the valence $d$ or $\bar d$ quark,
this matrix element does receive the contribution from the scalar
$f_0$ pole,
\begin{eqnarray} \label{eq:KKddme}
\langle K^+(p_2) K^-(p_3)|\bar dd|0\rangle^R
\equiv f^{K^+K^-}_d(s_{23})=\sum_{i}\frac{m_{{f_0}_i} \bar
f^d_{{f_0}_i}g^{{f_0}_i\to K^+K^-}}{m_{{f_0}_i}^2-s_{23}-i
m_{{f_0}_i}\Gamma_{{f_0}_i}},
\end{eqnarray}
which leads to
\begin{eqnarray}
 A_3 &=& {m_B^2-m_K^2\over
 m_b-m_s}F_0^{B_sK}(s_{23})f_d^{K^+K^-}(s_{23}) .
 \end{eqnarray}
For the equations $A_2$ and $A_4$, the contributions from
nonresonant could be parameterized as $F_{NR}$ and $f_d^{NR}$
respectively by using   $SU(3)$ symmetry  The formulae of $f_d^{NR}$
is expressed and discussed in detail in \cite{Cheng:2013dua}. After
calculation, we obtain
\begin{eqnarray}
  A_2 &=& (s_{12}-s_{23}) F_1^{B_sK}(s_{13}) F_{NR}(s_{13}),\\
  A_4 &=& {m_B^2-m_K^2\over
 m_b-m_s}F_0^{B_sK}(s_{13})f_d^{NR}(s_{13}).
 \end{eqnarray}
\subsection{$\overline B_s^0\to K^0 K^-\pi^+$  and $\overline B_s^0\to \overline K^0 K^+\pi^-$}
The factorizable amplitudes of the $\overline B_s^0 \to K^0
K^-\pi^+$ and $\overline B_s^0 \to \overline K^0 K^+\pi^-$ are given
as :
\begin{eqnarray} \label{eq:AK0Kmpip}
\langle K^0 K^-\pi^+|T_p|\overline B_s^0\rangle &=& \langle K^0\pi^+
|(\bar u b)_{V-A}|\overline B_s^0\rangle \langle K^-|(\bar s
u)_{V-A}|0\rangle
 \left[a_1 \delta_{pu}+a^p_4+a_{10}^p-r_\chi^K(a^p_6+a^p_8)\right]
 \nonumber\\
&& + \langle K^0|(\bar db)_{V-A}|\overline B_s^0\rangle \langle
K^-\pi^+|(\bar
sd)_{V-A}|0\rangle\Big[a_4^p-{1\over 2}a_{10}^p\Big] \nonumber \\
&& +\langle K^0|\bar db|\overline B_s^0\rangle \langle K^-\pi^+|\bar
dd|0\rangle
\Big[-2a_6^p+a_8^p\Big] \nonumber \\
&& +\langle K^0 K^-\pi^+|(\bar uu)_{V-A}|0\rangle \langle0|(\bar
sb)_{V-A}|\overline B_s^0\rangle\Big[a_2\delta_{pu}+a_3+a9\Big] \nonumber \\
&& +\langle K^0 K^-\pi^+|(\bar dd)_{V-A}|0\rangle \langle0|(\bar
sb)_{V-A}|\overline B_s^0\rangle\Big[a_3-{1\over 2}a_9\Big] \nonumber \\
&& +\langle K^0 K^-\pi^+|(\bar ss)_{V-A}|0\rangle \langle0|(\bar
sb)_{V-A}|\overline B_s^0\rangle\Big[a_3+a_4^p-{1\over 2}a_{9}-{1\over 2}a_{10}^p\Big] \nonumber \\
&& +\langle K^0 \pi^+ \pi^-|\bar s(1+\gamma_5)s|0\rangle
\langle0|\bar s\gamma_5b|\overline
B_s^0\rangle\Big[2a_6^p-a_8^p\Big],
\end{eqnarray}
\begin{eqnarray} \label{eq:AK0barKppim}
\langle \overline K^0 K^+\pi^-|T_p|\overline B_s^0\rangle &=&
\langle K^+\pi^- |(\bar db)_{V-A}|\overline B_s^0\rangle \langle
\overline K^0|(\bar sd)_{V-A}|0\rangle \left[a^p_4-{1\over 2}a_{10}^p-r_\chi^K(a^p_6-{1 \over 2}a^p_8)\right] \nonumber\\
&& + \langle K^+|(\bar ub)_{V-A}|\overline B_s^0\rangle
\langle \overline K^0\pi^-|(\bar su)_{V-A}|0\rangle\Big[a_1\delta_{pu}+a_4^p+a_{10}^p\Big] \nonumber \\
&& +\langle K^+|\bar ub|\overline B_s^0\rangle \langle \overline
K^0\pi^-|\bar su|0\rangle
\Big[-2a_6^p-2a_8^p\Big] \nonumber \\
&& +\langle \overline K^0 K^+\pi^-|(\bar uu)_{V-A}|0\rangle
\langle0|(\bar
sb)_{V-A}|\overline B_s^0\rangle\Big[a_2\delta_{pu}+a_3+a_9\Big] \nonumber \\
&& +\langle \overline K^0 K^+\pi^-|(\bar dd)_{V-A}|0\rangle
\langle0|(\bar
sb)_{V-A}|\overline B_s^0\rangle\Big[a_3-{1\over 2}a_9\Big] \nonumber \\
&& +\langle \overline K^0 K^+\pi^-|(\bar ss)_{V-A}|0\rangle
\langle0|(\bar
sb)_{V-A}|\overline B_s^0\rangle\Big[a_3+a_4^p-{1\over 2}a_{9}-{1\over 2}a_{10}^p\Big] \nonumber \\
&& +\langle \overline K^0 K^+\pi^-|\bar s(1+\gamma_5)s|0\rangle
\langle0|\bar s\gamma_5b|\overline
B_s^0\rangle\Big[2a_6^p-a_8^p\Big],
\end{eqnarray}
For the current-induced processes, the three-body matrix elements
$\langle K\pi |(\bar q b)_{V-A}|\overline B_s^0\rangle$ have the
similar expressions as Eqs.(\ref{eq:r&omega}) and
(\ref{eq:AHMChPT}). Furthermore, these processes also receive
resonant contributions, which is similar to Eq.(\ref{3formfactor})
except that the symbols of the final mesons are exchanged. For the
two-body matrix element $\langle K^-\pi^+|(\bar sd)_{V-A}|0\rangle$,
we note that
\begin{eqnarray}
\langle K^-(p_1)\pi^+(p_2)|(\bar sd)_{V-A}|0\rangle= (p_1-p_2)_\mu
F_1^{K\pi}(s_{12})
 + {m_K^2-m_\pi^2\over
 s_{12}}(p_1+p_2)_\mu\Big[-F_1^{K\pi}(s_{12})+F_0^{K\pi}(s_{12})\Big],
\end{eqnarray}
The resonant contributions are expressed by:
\begin{eqnarray} \label{eq:m.e.pole2}
\langle K^-(p_1)\pi^+(p_2)|(\bar sd)_{V-A}|0 \rangle^R &=& \sum_i
{g^{K^*_i\to K^-\pi^+}\over
s_{12}-m_{K^*_i}^2+im_{K^*_i}\Gamma_{K^*_i}}\sum_{\rm
pol}\varepsilon^*\cdot
(p_1-p_2)\langle K^*_i|(\bar sd)_{V-A}|0\rangle \nonumber \\
&-& \sum_i{g^{{K^*_{0i}}\to K^-\pi^+}\over s_{12}-
m_{K^*_{0i}}^2+im_{K^*_{0i}}\Gamma_{K^*_{0i}}}\langle K^*_{0i}|(\bar
sd)_{V-A}|0\rangle.
\end{eqnarray}
Hence, form factors $F_1^{K\pi}$ and $(-F_1^{K\pi}+F_0^{K\pi})$
receive the following resonant contributions
\begin{eqnarray} \label{eq:F1Kpi}
 (F^{K\pi}_{1}(s))^R &=& \sum_i{m_{K_i^*}f_{K_i^*}g^{K_i^*\to K\pi}\over
 m_{K_i^*}^2-s-im_{K_i^*}\Gamma_{K_i^*}}, \nonumber \\
 (-F^{K\pi}_1(s)+F^{K\pi}_0(s))^R &=& \sum_i{f_{K_{0i}^*}g^{K_{0i}^*\to K\pi}\over
 m_{K_{0i}^*}^2-s-im_{K_{0i}^*}\Gamma_{K_{0i}^*}}\,{s_{12}\over m_K^2-m_\pi^2}
 -\sum_i{m_{K_i^*}f_{K_i^*}g^{K_i^*\to K\pi}\over
 m_{K_i^*}^2-s-im_{K_i^*}\Gamma_{K_i^*}}{s_{12}\over m^2_{K_i^*}}.
\end{eqnarray}
As a result, the amplitude $\langle K^-\pi^+|(\bar
sd)_{V-A}|0\rangle$ $\langle
K^0|(\bar d b)_{V-A}|\overline
B_s^0\rangle$ has the expression
\begin{multline} \label{eq:KpiBpi}
\langle K^-(p_1)\pi^+(p_2)|(\bar s d)_{V-A}|0\rangle \langle
K^0(p_3)|(\bar d b)_{V-A}|\overline B_s^0\rangle\\
=F_1^{BsK}(s_{12})F_1^{K\pi}(s_{12})\left[s_{23}-s_{13}-{(m_B^2-m_K^2)(m_K^2-m_\pi^2)
\over s_{12}}\right] +
F_0^{BsK}(s_{12})F_0^{K\pi}(s_{12}){(m_B^2-m_K^2)(m_K^2-m_\pi^2)
\over s_{12}},
\end{multline}
where the momentum dependence of the weak form factor $F^{K\pi}(q^2)$ is parameterized as
\begin{eqnarray} \label{Kpi}
 F^{K\pi}(q^2)=\,{F^{K\pi}(0)\over 1-q^2/{\Lambda_\chi}^2+i\Gamma_R/{\Lambda_\chi}},
\end{eqnarray}
with $\Gamma_R= 200$ MeV \cite{Cheng:2002qu} being the width of the
relevant resonance and $\Lambda_{\chi}=0.83 \mathrm{GeV}$ being a
chiral symmetry breaking scale.

For the term $\langle K \pi|\bar s d|0\rangle$, it receives
contributions of both resonant and nonresonant, the expression of
which is shown as
\begin{eqnarray}
\langle K^-(p_1) \pi^+(p_2)|\bar s d|0\rangle
 = \frac{m_{{K^*_0}} \bar f_{{K^*_0}} g^{{K^*_0}\to K^-\pi^+}}{m_{{K^*_0}}^2-s_{12}-i
 m_{{K^*_0}}\Gamma_{{K^*_0}}}+\langle K^-(p_1) \pi^+(p_2)|\bar s
 d|0\rangle^{NR}.
\end{eqnarray}
In the above equation, the unknown two-body matrix elements of
scalar densities $\langle K\pi|\bar s q|0\rangle$ are related to
$\langle K^+K^-|\bar ss|0\rangle$ via SU(3) symmetry, e.g.
\begin{eqnarray}
\label{eq:Kpim.e.SU3} \langle K^-(p_1) \pi^+(p_2)|\bar
sd|0\rangle^{NR}=\langle K^+(p_1)K^-(p_2)|\bar
 ss|0\rangle^{NR}=f_s^{NR}(s_{12}),
\end{eqnarray}
with the expression of $f_s^{NR}$ given as
\begin{eqnarray}\label{fNR}
f_s^{NR}=\langle K^-(p_1) \pi^+(p_2)|\bar s
d|0\rangle^{NR}=\frac{m_K^2-m_\pi^2}{m_s-m_d}
(F_{NR}+\frac{2}{3}F'_{NR})+\sigma_{_{\rm NR}}
 e^{-\alpha s_{12}},
\end{eqnarray}
where $\sigma_{_{\rm NR}} = e^{i\pi/4} (3.36^{+1.12}_{-0.96})
\mathrm{GeV}$ is fixed from data of $\overline B^0 \to K_S K_S K_S$
\cite{Amhis:2012bh}. If we adopt this value directly, we will get
unexpected large branching fractions of $\overline B_s^0 \to
K_SK^\mp\pi^\pm $, which means that final states interaction and
$SU(3)$ symmetry violation may be important. Thus, we phonologically
introduce a factor $\beta=0.8\pm0.1$, which stands for the effects
of final states interaction and $SU(3)$ symmetry violation. While in
Ref.\cite{Cheng:2013dua}, a strong phase has been also introduced in
order to describe this effect.  As a result, we could obtain:
\begin{eqnarray}
\langle K^0|\bar db|\overline B_s^0\rangle \langle K^-\pi^+|\bar
sd|0\rangle = {m_{B_s}^2-m_K^2\over m_b-m_d}
F_0^{B_sK}(s_{12})\Big[\frac{m_{{K^*_0}} \bar f_{{K^*_0}}
g^{{K^*_0}\to K^-\pi^+}}{m_{{K^*_0}}^2-s_{12}-i
 m_{{K^*_0}}\Gamma_{{K^*_0}}}+\beta f_s^{NR}\Big].
\end{eqnarray}
\section{Numerical Results}\label{S3}
To proceed with the numerical calculations, we firstly specify the
parameters used in this work. For the CKM matrix elements, we use
the updated Wolfenstein parameters $A=0.823$, $\lambda=0.22457$,
$\bar \rho=0.1289$ and $\bar \eta=0.348$ \cite{CKMfitter}. The
corresponding CKM angles are $\sin2\beta = 0.689\pm0.019$ and
$\gamma=(69.7^{+1.3}_{-2.8})^\circ$. The form factors used in this
work are calculated within the covariant light-front quark model
\cite{Cheng:2009mu, Verma:2011yw}, which are summarized as follows
\begin{eqnarray}
&&V^{B_s\to \phi}(0)=0.23, A_0^{B_s\to\phi}(0)=0.31, A_1^{B_s\to \phi}(0)=0.25,
A_2^{B_s\to \phi}(0)=0.22,  \nonumber \\
&&V^{B_s\to K^*}(0)=0.23,A_0^{B_s\to K^*}(0)=0.25,A_1^{B_s\to K^*}(0)=0.19, A_2^{B_s\to K^*}(0)=0.16,\nonumber \\
&& F_{0}^{B_s \to K}(0)=0.24,F_{0}^{Bs \to
K_0^*(1430)}(0)=0.25, F_{0}^{Bs \to f_0^s}(0)=0.28.
\end{eqnarray}
The momentum dependence of form factors in the spacelike region can
be well parameterized and reproduced in the following
three-parameter form:
\begin{eqnarray}
F(q^2)=\frac{F(0)}{1-a(q^2/m_{B_s}^2)+b(q^2/m_{B_s}^2)^2}
\end{eqnarray}
where $F$ stands for the relevant form factors and parameters $a$
and $b$ have been given explicitly in ref.\cite{Verma:2011yw}.

In practical calculation, we shall assign the form factor error to
be $0.03$. For the strong coupling constants, most of them have been
determined from the measured partial width in
refs.\cite{Cheng:2007si, Cheng:2013dua}, which are shown as
\begin{eqnarray} \label{eq:g}
 && g^{\rho(770)\to\pi^+\pi^-}=6.0, \, g^{K^*(892)\to K^+\pi^-}=4.59,
\, g^{f_0(980)\to\pi^+\pi^-}=1.18\,{\rm GeV},\,
g^{K_0^*(1430)\to K^+\pi^-}=3.84\,{\rm GeV},\nonumber \\
&&g^{\phi\to K^+K^-}=-4.54,\,g^{f_0(980)\to
K^+K^-}=3.7~\mathrm{GeV},\,g^{f_0(1500)\to K^+K^-}=0.69
~\mathrm{GeV},\,g^{f_0(1710)\to K^+K^-}=1.6 ~\mathrm{GeV}.\nonumber\\
\end{eqnarray}
\begin{table}[htb!]
\begin{center}
\caption{Branching fractions (in units of $10^{-6}$) of resonant and
nonresonant (NR) contributions to $\overline B_s ^0 \to K_s^0
h^+h^{\prime -}$.}\label{tab:branching}
\begin{tabular}{l l | l l }
\hline \hline
 Decay mode~~~~~~~~~~~~~ & Theory~~~~~~~~~~~~~~~~~~~~~
& Decay mode & Theory \\
\hline $\overline B_s^0\to  K^0\pi^+\pi^-$
  \\
  \hline
$K^{*+}\pi^-$
  & $7.72^{+0.00+1.44+0.04}_{-0.00-1.27-0.04}$
& $K^0 f_0(980)$
  & $0.25^{+0.00+0.09+0.01}_{-0.00-0.07-0.00}$
  \\
$K^{*+}_0(1430)\pi^-$
  & $2.91^{+0.00+0.77+0.02}_{-0.00-0.67-0.02}$
& $K^0 f_0(1370)$
  & $0.25^{+0.00+0.07+0.00}_{-0.00-0.06-0.01}$
  \\
$K^0 \rho^0$
  & $0.53^{+0.00+0.14+0.01}_{-0.00-0.13-0.01}$
& NR
  & $2.90^{+0.68+0.37+0.05}_{-0.77-0.29-0.05}$ 
  \\
\hline Total
  & $12.58^{+0.49+2.42+0.10}_{-0.65-2.08-0.11}$ 
  &
  &
  \\ \hline\hline
  \\
 \hline
$\overline B_s^0\to  K^0K^+K^-$
  \\
  \hline
$\phi K^{0}$
  & $0.18^{+0.00+0.15+0.01}_{-0.00-0.08-0.01}$
& $f_0(980)K^{0}$
  & $0.20^{+0.00+0.16+0.00}_{-0.00-0.08-0.00}$
  \\
$f_0(1500)K^{0}$
  & $0.10^{+0.00+0.02+0.00}_{-0.00-0.02-0.00}$
& NR
  & $1.87^{+0.01+0.71+0.04}_{-0.01-0.59-0.05}$ 
  \\
\hline Total
  & $2.29^{+0.01+1.17+0.05}_{-0.01-0.78-0.05}$ 
    &
  & \\ \hline\hline\\
  \hline $\overline B_s^0\to   K^0 K^-\pi^+$
  \\
\hline $K^{*+}K^-$ &
$1.27^{+0.00+2.03+0.03}_{-0.00-0.73-0.04}$
& $\overline K^{*0}K^0$ &
$2.27^{+0.00+0.60+0.01}_{-0.00-0.53-0.01}$
  \\
$K_0^{*+}(1430)K^-$
  & $0.89^{+0.00+1.43+0.02}_{-0.00-0.51-0.02}$
& $\overline K_0^{*0}(1430)K^0$
  & $16.63^{+0.00+5.12+0.02}_{-0.00-4.32-0.03}$
  \\
NR
  & $12.89^{+0.28+13.17+0.05}_{-0.36-~6.56-0.05}$ 
& &
\\
\hline Total & $34.24^{+0.23+21.11+0.09}_{-0.35-11.93-0.08}$ 
& &
\\ \hline \hline\\
\hline $\overline B_s^0\to  \overline K^0 K^+\pi^-$
\\
\hline $K^{*0}\overline K^0$
&$0.73^{+0.00+1.70+0.00}_{-0.00-0.53-0.00}$
& $K^{*-}K^+$ & $2.27^{+0.00+0.60+0.00}_{-0.00-0.54-0.00}$
  \\
$K_0^{*0}(1430)\overline K^0$ &
$0.51^{+0.00+1.20+0.00}_{-0.00-0.37-0.00}$
& $K_0^{*-}(1430)K^+$ & $15.47^{+0.00+4.55+0.00}_{-0.00-3.89-0.00}$
\\
NR & $12.29^{+0.25+12.58+0.02}_{-0.32-~6.29-0.02}$ 
&
&
\\
\hline Total & $33.71^{+0.15+20.93+0.01}_{-0.19-11.95-0.01}$ 
& &
  \\ \hline \hline
\end{tabular}
\end{center}
\end{table}
For the running quark masses we shall use
\cite{Beringer:1900zz,Xing}
 \begin{eqnarray} \label{eq:quarkmass}
 && m_b(m_b)=4.2\,{\rm GeV}, \qquad~~~~ m_b(2.1\,{\rm GeV})=4.94\,{\rm
 GeV}, \qquad m_b(1\,{\rm GeV})=6.34\,{\rm
 GeV}, \nonumber \\
 && m_c(m_b)=0.91\,{\rm GeV}, \qquad~~~ m_c(2.1\,{\rm GeV})=1.06\,{\rm  GeV},
 \qquad m_c(1\,{\rm GeV})=1.32\,{\rm
 GeV}, \nonumber \\
 && m_s(2.1\,{\rm GeV})=95\,{\rm MeV}, \quad~ m_s(1\,{\rm GeV})=118\,{\rm
 MeV}, \nonumber\\
 && m_d(2.1\,{\rm GeV})=5.0\,{\rm  MeV}, \quad~ m_u(2.1\,{\rm GeV})=2.2\,{\rm
 MeV}.
\end{eqnarray}

With above parameters and formulas in Sec.\ref{S2}, we calculated
the branching fractions of resonant and nonresonant contributions to
the decay modes concerned and  presented  them in
Table.\ref{tab:branching}. The theoretical errors are from the
uncertainties in (i) the parameter $\alpha_{_{\rm NR}}$ which
governs the momentum dependence of the nonresonant amplitude, (ii)
the strange quark mass $m_s$, the form factors, the nonresonant
parameter $\sigma_{_{\rm NR}}$ and $SU(3)$ asymmetry violation
parameter $\beta$, and (iii) the unitarity angle $\gamma$.

From Table. \ref{tab:branching} we see that the decay $\overline
B_s^0\to  K^0\pi^+\pi^-$ is tree dominated and its main contribution
arises from the  $K^{*+}$ meson, while the nonresonant contribution
is less important. Compared with experimental data, the calculated
branching fraction agrees well with the recent LHCb measurement. As
for  $\overline B_s^0\to  K^0K^+K^-$, although it receives the
color-suppressed tree contribution, it is dominated by transition $b
\to d \bar qq $. Consequently, it has a small branching fraction
$(2.29^{+0.01+1.17+0.05}_{-0.01-0.78-0.05})\times 10^{-6}$, which is
much smaller than that of $\overline B_s^0\to K^0\pi^+\pi^-$. Note
that this decay is governed by the nonresonant background dominated
by $\sigma_{\rm NR}$. Hence this decay mode could be an ideal plat
for constraining  the unknown parameter $\sigma_{\rm NR}$ in turn.
Experimentally, however, no significant evidence of this decay mode
has been obtained, and its branching fraction is described in
$(0.2-3.4) \times 10^{-6}$ at $90\%$ confidence level (CL) based on
the CL inferences in Ref. \cite{Feldman:1997qc}. Obviously, the
result we predicted is falling into the experimental range. We hope
this decay will be measured precisely in the current LHCb
experiment. The results of above two decay modes  also confirm the
conclusion that nonresonant decays play a prominent role in the
penguin-dominated three-body $B$ meson decays in Ref.
\cite{Cheng:2007si}.

For the decay $\overline B_s^0 \to K^0K^-\pi^+$, the current-induced
process with a $K^-$ emission is tree dominated, while the
transition processes  $\langle \overline B_s^0 \to K^0\rangle
\times\langle 0 \to  K^-\pi^+\rangle$ are induced by penguin
operators.  On the contrary,  the current-induced process of  decay
$\overline B_s^0 \to \overline K^0K^+\pi^-$ with a neutral kaon
emission is induced by penguin,  and the  transition processes
receive the effects not only from tree but from penguin operators.
In these two decays, the nonresonant contributions arise dominantly
from the transition process via the scalar density $\langle
K\pi|\bar s q|0\rangle$, and slightly from the current-induced
process. Thus, the nonresonant contributions are sensitive to the
matrix elements of scalar densities $f_s^{\rm NR}$, as shown in
Table.\ref{tab:branching}. For the resonant contributions, both of
them are dominated by the  scalar particles $K_0^*(1430)$.
Considering the parameter $\beta$ standing for effects of the
$SU(3)$ symmetry violation and the final states rescattering,  the
sum of two branching fractions is
$(67.95^{+0.38+42.04+0.09}_{-0.54-23.88-0.08}) \times 10^{-6}$,
which could accommodate  data of the recent LHC measurement well. We
hope these two decays could be measured individually in the future
experiment.

In QCD calculations based on a heavy quark expansion, one faces
uncertainties arising from power corrections such as annihilation
and hard-scattering contributions. For example, in QCD factorization
\cite{BBNS}, there are large theoretical uncertainties related to
the modelling of power corrections corresponding to weak
annihilation effects and the chirally enhanced power corrections to
hard spectator scattering. Even for two-body $B$ decays, power
corrections are of order $(10-20)\%$ for tree-dominated modes, but
they are usually bigger than the central values for
penguin-dominated decays. Needless to say, $1/m_b$ power corrections
for three-body decays may well be larger. However, in the current
work we use the phenomenological factorization model rather than in
the established theories based on a heavy quark expansion.
Consequently, uncertainties due to power corrections, at this stage,
are not included in our calculations, by assumption. In view of such
shortcomings we must emphasize that the additional errors due to
such model dependent assumptions may be sizable.

In this work, the $CP$ asymmetries of these four decays are also
calculated, and the results are summarized in Table.\ref{tab:CP}. We
see from the table that the decay $\overline B_s^0 \to K^0 K^+K^-$
has large $CP$ asymmetries with and without resonant contributions.
Note that the two asymmetries have the same sign, as this decay is
dominated by the nonresonant background, which can also be read from
Table.\ref{tab:branching}. For other three decays, the sizable
resonant contributions may affect the $CP$ asymmetries by taking
large strong phases. In fact, the strong phases could arise from the
effective Wilson coefficients, the Breit-Wigner formalism for
resonances and the penguin matrix elements of scalar densities.
Besides, the final states interactions may take new phases, which
cannot be calculated directly up to now. Although the $CP$
asymmetries of  $B \to K K K \,, KK\pi$ \cite{LHCb:Kppippim,
LHCb:pippippim} have been measured in LHCb recently, the $CP$
asymmetries of three-body of $B_s^0$ have not been explored till
now. The $CP$ asymmetries of these four decays are hoped to be
measured in the current LHCb experiment or Super-b in the future,
and they might be helpful to test the factorization approach in
$B_s^0$ meson three-body decays.

\begin{table}[t]
\begin{center}
\caption{Direct $CP$ asymmetries (in \%) for decay modes of
$\overline B_s^0$ decays.} \label{tab:CP}
\begin{tabular}{l  c  c }
\hline\hline
 Final state~~ & Total  & Nonresonant  \\ \hline
$\overline B_s^0\to  K^0\pi^+\pi^-$
&$10.4^{+0.4+0.4+0.1}_{-0.5-0.7-0.1}$
&$23.6^{+1.6+0.8+0.2}_{-1.6-1.6-0.2}$\\
$\overline B_s^0\to  K^0K^+K^-$
&$-16.6^{+0.1+0.6+0.1}_{-0.2-0.5-0.1}$
&$-19.4^{+0.0+0.1+0.2}_{-0.0-0.1-0.2}$\\
$\overline B_s^0\to   K^0 K^-\pi^+$
&$-1.8^{+0.5+0.5+0.0}_{-0.5-0.6-0.1}$
&$-2.2^{+0.5+1.1+0.1}_{-0.4-0.9-0.1}$\\
$\overline B_s^0\to  \overline K^0 K^+\pi^-$
&$0.1^{+0.0+0.1+0.0}_{-0.0-0.3-0.0}$
&$0.7^{+0.0+0.0+0.0}_{-0.0-0.0-0.0}$
  \\
\hline
\end{tabular}
\end{center}
\end{table}

\section{Summary}\label{S4}
Recently, LHCb collaboration published their first measurements of
charmless three-body decays of $B_s^0$ meson, corresponding to an
integrated luminosity of $1.0~\mathrm{fb}^{-1}$ recorded at a
centre-of-mass energy of 7 TeV. Motivated by this, we calculated the
branching fractions of $\overline B_s ^0 \to K^0 \pi^+\pi^-$,
$\overline B_s ^0 \to K^0 K^+K^-$, $\overline B_s ^0 \to K^0
\pi^+K^-$ and $\overline B_s ^0 \to \overline K^0 K^+\pi^-$ decay
modes within the factorization approach, which is generalized by
Cheng {\it et al}.  Both nonresonant contributions and resonant
contributions have been studied in detail. For the decays $\overline
B_s ^0 \to K^0 \pi^+\pi^-$ and $\overline B_s ^0 \to K^0 K^+K^-$,
our results agree well with experimental data. Especially, the
former mode is dominated by the $K^*$ and $K_0^*(1430)$ poles, while
the latter is dominated by the nonresonant contribution. By adding
the effects of the flavor $SU(3)$ symmetry violation, the sum of
branching fractions of $\overline B_s ^0 \to K^0 \pi^+K^-$ and
$\overline B_s ^0 \to \overline K^0 K^+\pi^-$ could accommodate the
data. It should be emphasized  that the branching fractions are very
sensitive to the scalar density $\langle K\pi| \bar s q|0\rangle$.
We hope these branching fractions could be measured individually in
the experiments so as to test the factorization approach in
three-body decays of $\overline B^0_s$ mesons.  Moreover, the direct
$CP$ asymmetries of these decays have been also explored, and the
sizable results could be measured in the running LHCb experiment and
Super-b factory in the future.

\section*{Acknowledgments}
Y. Li thanks Hai-Yang Cheng and Chun-Khiang Chua for valuable
discussions and comments. This work is supported by the National
Science Foundation (Grants No. 11175151 and No. 11235005), and the
Program for New Century Excellent Talents in University (NCET) by
Ministry of Education of P. R. China (Grant No. NCET-13-0991).
\section*{Note added}
When this paper is being prepared, Hai-Yang Cheng and Chun-Khiang
Chua posted their paper to the e-print archiv \cite{Cheng:2014uga}.
The same decays have been studied in that work,  and most of our
results agree with theirs after considering the differences of
parameters (form factors). In \cite{Cheng:2014uga}, much attention
is paid to the $U$-spin asymmetry, while in this work we paid much
attention to disentangle the resonant and nonresonant contributions.
Moreover, in dealing with the flavor $SU(3)$ symmetry violation of
$\langle K\pi|0\rangle$, different approaches are adopted.


\end{document}